\documentclass[conference]{IEEEtran}
\IEEEoverridecommandlockouts
\usepackage{cite}
\usepackage{amsmath,amssymb,amsfonts}
\usepackage{algorithmic}
\usepackage{graphicx}
\usepackage{textcomp}
\usepackage{subcaption}
\usepackage{graphicx}
\usepackage{xcolor}

\def\BibTeX{{\rm B\kern-.05em{\sc i\kern-.025em b}\kern-.08em
    T\kern-.1667em\lower.7ex\hbox{E}\kern-.125emX}}
\begin{document}

\title{Analyzing crosstalk error in the NISQ era
\thanks{This work is funded by the QuantUM Initiative of the Region Occitanie, France, University of Montpellier, France, and IBM Montpellier, France.}
}

\author{\IEEEauthorblockN{Siyuan Niu}
\IEEEauthorblockA{\textit{LIRMM, University of Montpellier} \\
34095 Montpellier, France \\
siyuan.niu@lirmm.fr}
\and
\IEEEauthorblockN{Aida Todri-Sanial}
\IEEEauthorblockA{\textit{LIRMM, University of Montpellier, CNRS} \\
34095 Montpellier, France \\
aida.todri@lirmm.fr}
}

\maketitle

\begin{abstract}
Noisy Intermediate-Scale Quantum (NISQ) hardware has unavoidable noises, and crosstalk error is a significant error source. When multiple quantum operations are executed simultaneously, the quantum state can be corrupted due to the crosstalk between gates during simultaneous operations, decreasing the circuit fidelity. In this work, we first report on several protocols for characterizing crosstalk. Then, we discuss different crosstalk mitigation methods from the hardware and software perspectives. Finally, we perform crosstalk injection experiments on the IBM quantum device and demonstrate the fidelity improvement with the crosstalk mitigation method. 

%

\end{abstract}


\section{Introduction}

Quantum Computing is expected to solve specific classical intractable problems. Several companies such as IBM, Google, and Rigetti have released their quantum chips with 65, 72, and 31 qubits, respectively. However, these chips are qualified as Noisy Intermediate-Scale Quantum (NISQ) hardware, which have less than one hundred qubits and suffer from unavoidable noises. 

Crosstalk is one of the major noise sources not only in superconducting but also in trapped-ion devices. It can corrupt the qubit state when multiple quantum operations are executed simultaneously. Crosstalk has a significant impact on quantum gate error. For instance,~\cite{murali2020software} shows an increase of \texttt{CNOT} error up to 11 times caused by crosstalk. Different protocols were proposed in~\cite{gambetta2012characterization,erhard2019characterizing,bialczak2010quantum} to detect and characterize crosstalk in quantum devices. After assessing crosstalk, there are two types of methods to mitigating it. The first one is based on hardware strategies, such as tunable coupling~\cite{mundada2019suppression}, or frequency allocation~\cite{li2020towards}. From the software perspective, crosstalk has been mitigated by changing simultaneous \texttt{CNOT} operations with high crosstalk to execute separately while trading off the increase of the decoherence time as in~\cite{murali2020software}.

In this work, we first introduce how to characterize a quantum device's crosstalk using protocol Simultaneous Randomized Benchmarking (SRB)~\cite{gambetta2012characterization}. Second, we present various state-of-the-art crosstalk mitigation methods. Third, we inject crosstalk on IBM quantum devices to show its impact on output fidelity. Finally, we evaluate the crosstalk mitigation method using several benchmarks to demonstrate the fidelity improvement.
%
\section{Methods}

\subsection{Crosstalk characterization using SRB}

To characterize the crosstalk effect of a quantum device, we choose the most commonly used protocol -- Simultaneous Randomized Benchmarking (SRB)~\cite{gambetta2012characterization} based on its ability to quantify the impact of parallel instructions, such as the crosstalk effect of one quantum gate to another when they are executed at the same time. To characterize the crosstalk effect between gate $g_i$ and $g_j$, we first perform Randomized Benchmarking (RB) on both gates separately and obtain their independent error rate $\mathcal{E}(g_i)$ and $\mathcal{E}(g_j)$. Then, applying SRB on both gates yields the correlated error rate $\mathcal{E}(g_i|g_j)$ and $\mathcal{E}(g_j|g_i)$ for simultaneous executions. If there exists crosstalk between them, the relation between independent and correlated errors should comply with $\mathcal{E}(g_i|g_j) > \mathcal{E}(g_i)$ or $\mathcal{E}(g_j|g_i) > \mathcal{E}(g_j)$. Previous work~\cite{murali2020software} has demonstrated that crosstalk is significant due to simultaneous \texttt{CNOT} executions, hence, in this work, we only focus on characterizing the crosstalk effect between \texttt{CNOT} pairs. 

We use the ratio of correlated error to independent error $r(g_i|g_j) = \mathcal{E}(g_i|g_j) / \mathcal{E}(g_i)$ as the indicator of the crosstalk effect of the \texttt{CNOT} pairs. We choose IBM Q 7 Casablanca as an example to show the crosstalk effect on this device. As \texttt{CNOT} errors vary over each calibration, we characterize the crosstalk effect twice respectively on each possible simultaneous \texttt{CNOT} pair of IBM Q Casablanca.

\begin{figure}[h]

	\centering
	\includegraphics[scale=0.5]{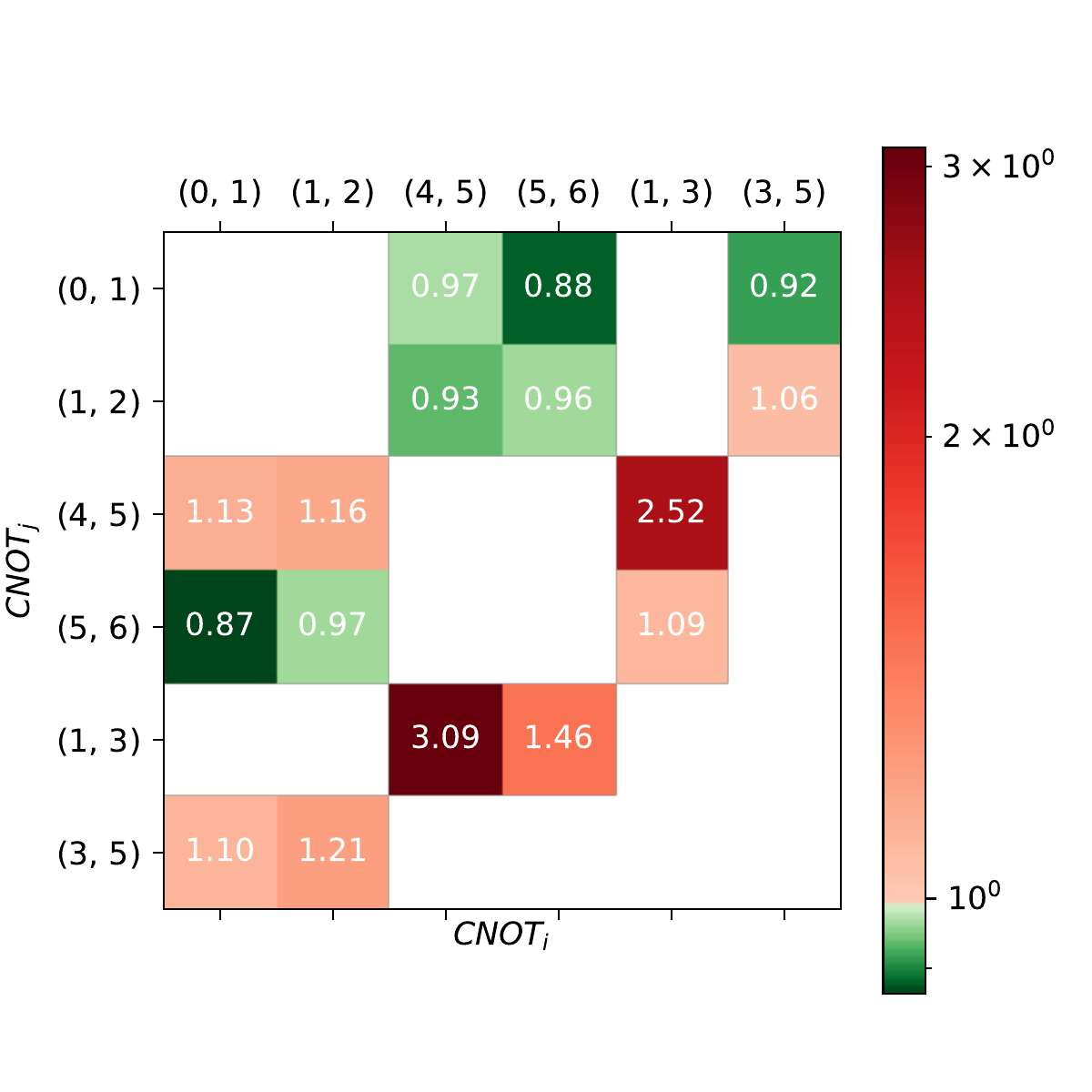}

	\caption{The SRB result of IBM Q 7 Casablanca. The number represents the ratio of correlated error to independent error. Note that the SRB experiment is performed on \texttt{CNOTs} that can be executed in parallel, which means they do not share the same qubits. There are no SRB experiments on blank spots due to the sharing qubits.}
	\label{fig:SRB}
\end{figure}

The results of average crosstalk data are presented in Fig.~\ref{fig:SRB}. The crosstalk effect remains stable regardless of the variation of gate errors across days. For each experiment, there are 150 Cliffords in the sequence and each sequence is repeated five times due to the expensive runtime cost. The crosstalk effect is not as severe in this device as in other devices such as IBM Q 20 Poughkeepsie, where the gate error grows up to 11 times caused by crosstalk~\cite{murali2020software}. However, the error rate is still amplified up to 3 times in IBM Q Casablanca. Sometimes ratios are less than one, which is probably due to the limits of crosstalk metric in SRB protocol~\cite{mckay2020correlated}.

\subsection{Crosstalk mitigation}
After characterizing the crosstalk effect, the next challenge is how to mitigate it. The hardware-oriented crosstalk mitigation strategies have several physical constraints. For example, the tunable coupling~\cite{mundada2019suppression} is
only applicable for tunable coupling superconducting devices, whereas the frequency allocation~\cite{li2020towards} is mainly designed for fixed-frequency transmon qubit devices. Despite hardware improvement approaches to reduce crosstalk, there are also software methods to address crosstalk. Here, we report on hardware-agnostic software crosstalk mitigation approaches.

An intuitive approach for software crosstalk mitigation is to make simultaneous \texttt{CNOTs} execute serially to avoid simultaneous high crosstalk \texttt{CNOTs}. However, serial instructions can increase the circuit depth and cause decoherence errors. Therefore, a scheduler is required to trade-off between them. 
The state of the art~\cite{murali2020software} proposes a crosstalk-adaptive scheduler (labeled as XtalkSched) based on SMT optimization that re-schedules a quantum intermediate representation (IR), taking into account both crosstalk and decoherence error. It inserts barriers between the simultaneous \texttt{CNOTs} with a significant crosstalk effect to avoid parallel instructions while accounting for the qubit coherence time.

\section{Evaluation}
\subsection{Methodology}

\begin{figure}[t]
	
	\centering
	\includegraphics[scale=0.6]{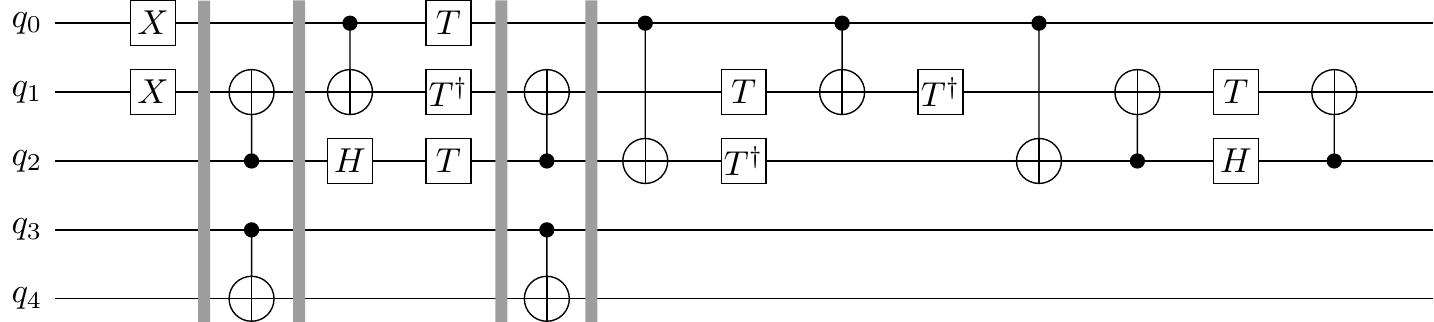}
	
	\caption{Crosstalk injection of CSWAP circuit. We introduce two other qubits ($q_3$, $q_4$) and apply \texttt{CNOT} operations on them. Barriers are inserted to make sure the \texttt{CNOTs} are executed simultaneously. Note that, crosstalk injection are performed on \texttt{CNOT} pairs with strong crossalk error.}
	\label{fig:cswap}
\end{figure}

\textbf{Benchmarks.} We first use a 3-qubit CSWAP gate circuit and inject a different number of simultaneous \texttt{CNOTs} to show the impact of crosstalk on the output fidelity of a quantum circuit. An example of demonstrating crosstalk injection by enabling two simultaneous \texttt{CNOTs} is shown in Fig.~\ref{fig:cswap}. Then, we examine benchmarks collected from the previous works~\cite{murali2020software,li2020qasmbench}, including SWAP circuits, Bell state circuit, etc., to show the performance of XtalkSched. For each benchmark, we select its mapping to ensure it includes at least one pair of \texttt{CNOTs} with a high crosstalk effect.

\textbf{Algorithm configuration.} 
For crosstalk characterization of IBM Q Casablanca, we set the crosstalk threshold to 2, and $w$ used in XtalkSched is set to 0.5 to trade-off circuit depth and crosstalk mitigation. The Qiskit version is 0.23.6.

\textbf{Comparison.} We compare XtalkSched~\cite{murali2020software} with ParSched, which is the current scheduler used in Qiskit to make instructions execute in parallel without considering the crosstalk.

\subsection{Experimental results}
\begin{figure}
	\centering
	\includegraphics[scale=0.6]{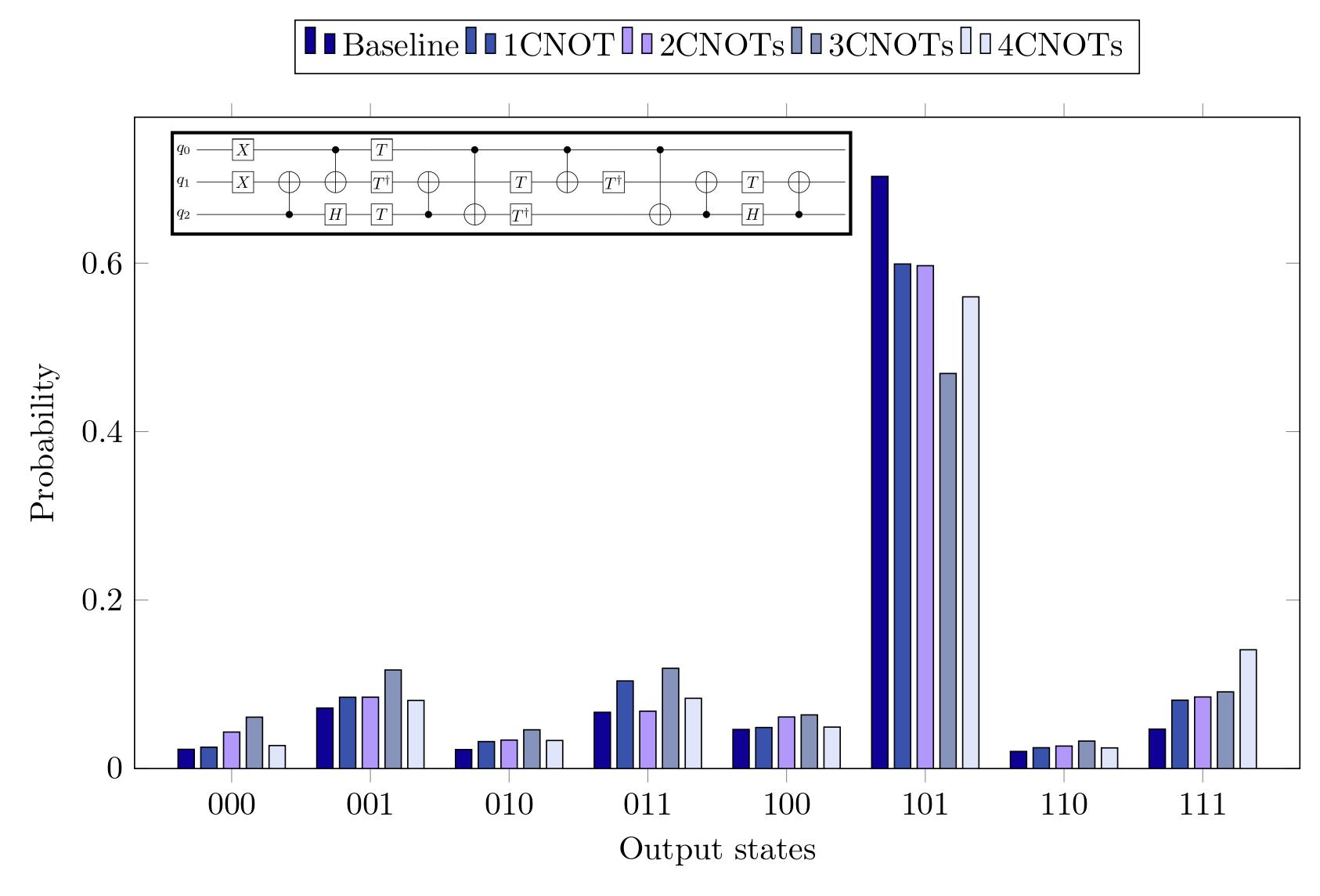}
	\caption{The output state probability distribution of CSWAP gate circuit. The right output state should be '101'. We inject a different number of simultaneous \texttt{CNOTs} to this circuit. The baseline circuit has zero simultaneous \texttt{CNOT}.}
	\label{fig:injection}
\end{figure}

The experimental results of crosstalk injection are shown in Fig.~\ref{fig:injection}. The probability of obtaining the right output state is reduced with the increase of simultaneous \texttt{CNOTs} due to the injected crosstalk error. In the worst case, the fidelity is decreased by 33.3\% compared to the non-crosstalk case. 

\begin{figure}[h]
	\begin{subfigure}{0.45\columnwidth}
		\centering
		\caption{}
		\includegraphics[scale=0.6]{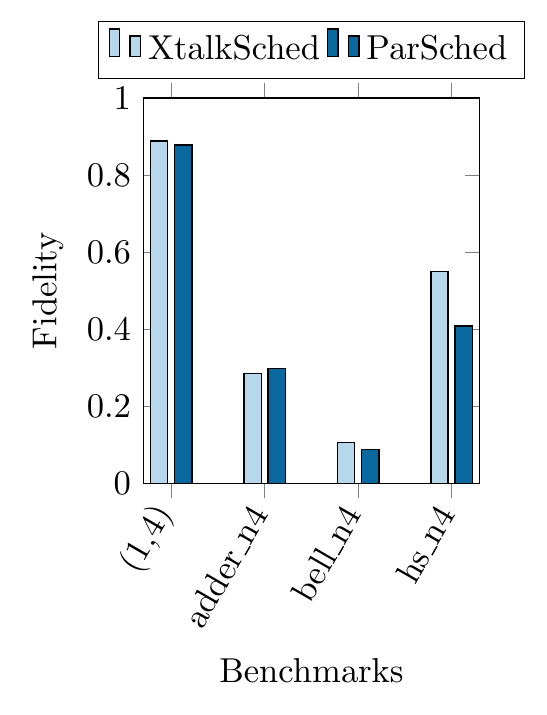}
		\label{fig:SRB1}
	\end{subfigure}
	\hspace{1mm}
	\begin{subfigure}{0.45\columnwidth}
		\centering
		\caption{}
		\includegraphics[scale=0.6]{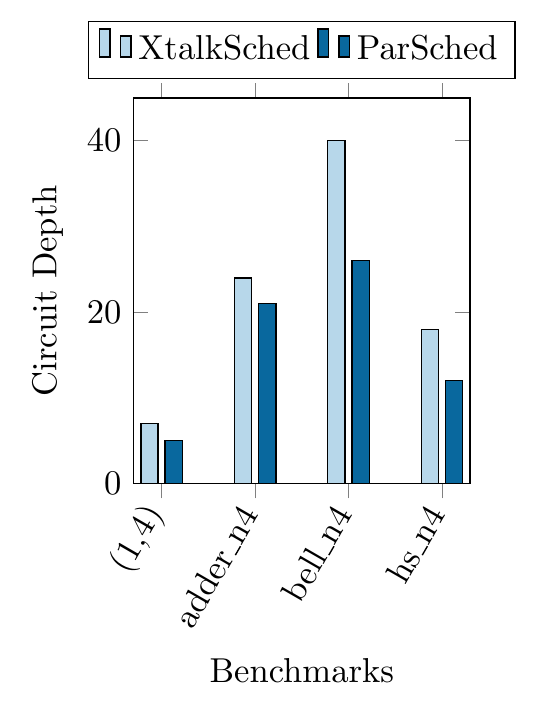}
		\label{fig:SRB2}
	\end{subfigure}
	
	\caption{(a) Fidelity. (b) Circuit depth. Note that (1,4) represents the SWAP circuit which aims to realize a connection between $q_0$ and $q_4$ through SWAP operations.}
	\label{fig:result}
\end{figure}

The comparison between XtalkSched with ParSched in terms of output fidelity and circuit depth is shown in Fig.~\ref{fig:result}. The fidelity is improved by 9.4\%, whereas the circuit depth increased by 39.1\%. Although the fidelity is enhanced because of XtalkSched, the circuit depth increase is not negligible. This raises the need for more effective and practical crosstalk mitigation methods.

\section{Conclusion}
Crosstalk is a non-negligible error source in quantum computers. This work examines the crosstalk characterization protocol based on simultaneous randomized benchmarking, and we report on the crosstalk effects on the IBM quantum device. We also evaluate the crosstalk-adaptive scheduler and illustrate its impact on output fidelity and circuit depth. 



\bibliography{bibliography}{}
\bibliographystyle{plain}

\end{document}